\documentclass{aa}
\usepackage{times,graphics,psfig}
\def\rgb{RGB\,J0044+193 }
\begin{document}
\thesaurus{03(11.01.2; 11.09.1 RGB J0044+193; 13.18.1; 13.25.2)}
\title{An ASCA observation of the radio-loud narrow-line \\
Seyfert 1 galaxy RGB J0044+193}
\author{J. Siebert\inst{1,2}
\and K.M. Leighly\inst{3}
\and S.A. Laurent--Muehleisen\inst{4} 
\and W. Brinkmann\inst{1}
\and Th. Boller\inst{1}
\and M. Matsuoka\inst{2}}
\mail{J. Siebert, jos@mpe.mpg.de}
\institute{Max--Planck--Institut f\"ur extraterrestrische Physik,
 Giessenbachstrasse, 85740 Garching, Germany
\and
 Institute of Physical and Chemical Research (RIKEN), 2-1 Hirosawa, Wako,
 Saitama 351-01, Japan
\and 
 Columbia Astrophysics Laboratory, Columbia University, New York, NY 10027, USA
\and
 IGPP/LLNL, 7000 East Av., Livermore, CA 94550, USA}
\titlerunning{ASCA observation of RGB\,J0044+193}
\authorrunning{Siebert et al.}
\date{Received 3 November 1998 / Accepted 24 February 1999}
\maketitle
\begin{abstract}
Narrow-line Seyfert 1 galaxies are generally found to be radio-quiet and there
was only one radio-loud object known so far (PKS 0558-504). Here we present 
the results of a 50 ksec ASCA observation of the recently discovered second
radio-loud NLS1 galaxy RGB\,J0044+193. The X-ray data are complemented by
radio observations and a new optical spectrum for this source.

We find evidence for variable radio emission and an inverted radio spectrum
of RGB\,J0044+193. The optical continuum turned out to be extremely blue.
This may either indicate additional line emission, for example from 
Fe\,{\footnotesize I}, or scattering of a blue intrinsic continuum.   
The X-ray spectrum shows a clear break around 1.8 keV. Above this energy the 
spectrum is characterized by a power law with a photon index of $\Gamma\approx 
2.1$. For energies below 2 keV the spectrum is much softer and it can either 
be modeled with a steeper power law ($\Gamma\approx2.7$) or a blackbody 
component with a temperature around 0.2 keV. The X-ray count rate of the 
source decreased by a factor of two within one
day and there is evidence for low amplitude variability on much shorter
time scales. Given its average 2--10 keV X-ray luminosity of $(1.35\pm0.05)
\times 10^{44}$ erg s$^{-1}$, \object{\rgb} is significantly more variable 
than a typical broad line Seyfert 1 galaxy of comparable X-ray luminosity, 
but consistent with the bulk of NLS1 galaxies. 

The spectral as well as the variability properties of \rgb
are indistinguishable from those of radio-quiet NLS1s. 
In particular, we find no evidence for a flat X-ray component due to inverse 
Compton emission related to the putative non-thermal radio emission from
RGB\,J0044+193. We argue, however, that this does not rule out
a pole-on orientation for RGB\,J0044+193. 
\keywords{Galaxies: active -- individual:RGB\,J0044+193; X-rays: galaxies -- 
Radio continum: galaxies.}

\end{abstract}

\section{Introduction}

The radio-loud X-ray source \rgb (R.A. = 00:44:59.12 ; Dec. = +19:21:40.8; 
J2000.0) was serendipitously identified as a NLS1 in the course of a large 
effort to establish a new, well-defined and large sample of X-ray selected 
BL Lac objects (Laurent-Muehleisen et al. \cite{laurent98}). The basic sample 
resulted from a cross-correlation of the ROSAT All-Sky Survey and the Green 
Bank 5GHz radio survey (87GB; Condon et al. \cite{condon}). 
This so-called RGB-sample contains more than 2100 radio-loud X-ray sources 
and was completely followed-up with the VLA to obtain accurate radio positions
and radio core fluxes (Brinkmann et al. \cite{brinkmann95}, \cite{brinkmann97};
Laurent-Muehleisen et al. \cite{laurent97}).

In X-rays NLS1 generally exhibit rapid variability and steep X-ray spectra 
(e.g. Boller et al. \cite{boller96}; Brandt et al. \cite{brandt}; 
Leighly \cite{leighly98}, \cite{leighly99}). Whether or not the X-ray 
and optical properties 
of NLS1 galaxies are affected by orientation is currently the subject of much 
debate (e.g. Osterbrock \& Pogge \cite{osterbrock}; Brandt et al. 
\cite{brandt}; Boller et al. \cite{boller96}, \cite{boller97}). Originally it 
was suggested that NLS1 are the pole-on versions of regular Seyfert 1 galaxies.
This naturally explains the narrower permitted emission lines, if the emission
line clouds were confined to a disk perpendicular to the symmetry axis. 
Orientation dependent emission line properties have also been discussed for 
other object classes (e.g. Boroson \cite{boroson}; Baker \& Hunstead 
\cite{baker}). The high frequency of soft X-ray excesses seen in NLS1 (e.g. 
Leighly \cite{leighly99}) also might argue for a face-on orientation, at least 
in the case of geometrically thick accretion discs (Madau \cite{madau}). 
Further arguments in favor of pole-on geometries are the strong Fe II emission,
if it originates in the accretion disc (e.g. Kwan et al. 1995) and the low 
frequency of warm absorbers in NLS1 (Leighly \cite{leighly99}). Finally, the 
observed rapid variability as well as absorption features near 1 keV in three 
NLS1 might be explained by relativistic outflows seen close to the line of 
sight (Boller et al. \cite{boller96}; Leighly et al. \cite{leighly97}). 

On the other hand, the orientation model would require the hard X-ray spectral
index to be orientation dependent, since it is much flatter in broad-line 
Seyferts than in NLS1. However, current models for the X-ray emission processes
predict the opposite behavior (e.g. Haardt \& Maraschi \cite{haardt}). 
Further, the extreme variability in IRAS 13224-3809 can be explained by
relativistic effects connected with a hot spot at the inner edge of an 
accretion disc seen under a large inclination angle. Finally, the presence 
of excess absorption by cold
material in the NLS1 Mrk 507 (Iwasawa et al. \cite{iwasawa}) and the 
observation of ionization cones in Mrk 766 (Wilson et al. \cite{wilson}) argue
against a general pole-on orientation for NLS1.    

Instead, it has been proposed that NLS1 galaxies are objects with smaller 
black hole masses and accretion rates closer to the Eddington limit compared 
to broad-line Seyferts. This scenario naturally explains many of their 
peculiar optical and X-ray properties (Pounds et al. \cite{pounds}; 
Leighly \cite{leighly99} and references therein).  

Radio-loud NLS1 appear to be a rare class of AGN. Apart from \rgb there is
only one object known, namely \object{PKS 0558-504}. The importance of 
radio-loud NLS1 galaxies lies in the fact that the presence of radio emission 
and an associated relativistic jet might give us an independent handle on the 
orientation issue. Hence, the main intention of our ASCA observation was to 
compare the X-ray properties of the radio-loud NLS1 \rgb to radio-quiet NLS1 
and to investigate, whether any differences in the X-ray spectrum can be 
attributed to the radio-loud nature of this object.

\section{The optical spectrum of \rgb}

\begin{figure}
\resizebox{\hsize}{!}{\includegraphics{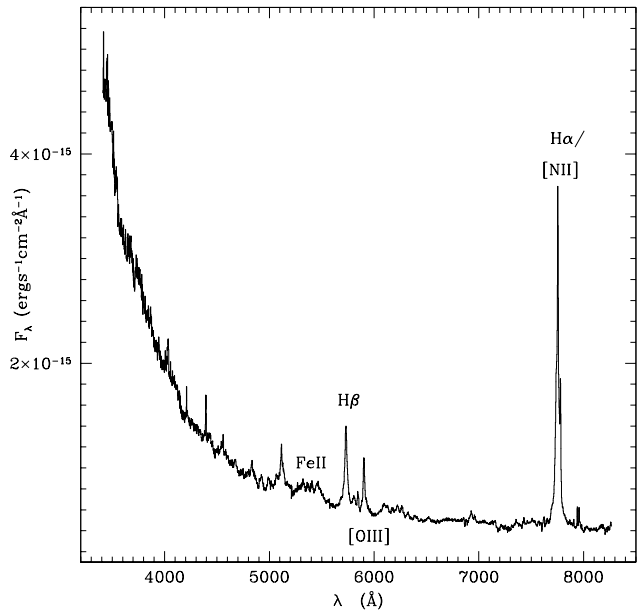}}
\caption{The observed broad band (3410 and 8270 {\AA}) optical spectrum of 
RGB\,J0044+193, taken with the Kast double spectrograph at Lick 
observatory. For clarity, only the emission lines discussed in the text are 
labeled. Note the extremely blue continuum below 5000{\AA}.}
\label{optb}
\end{figure}

\begin{figure}
\psfig{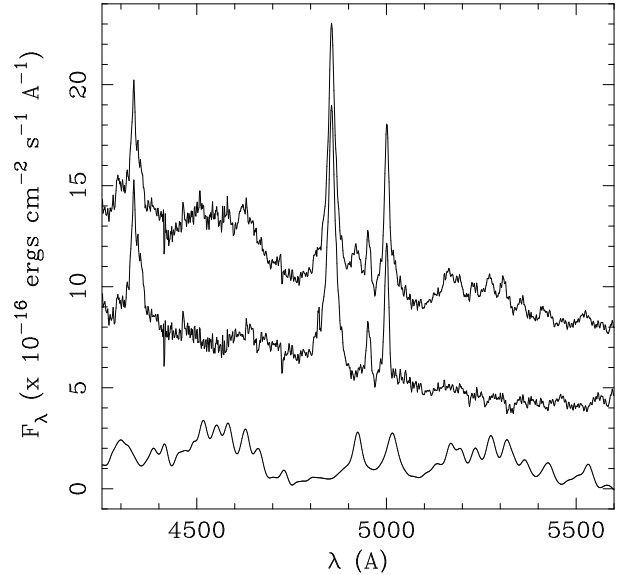}
\caption{To illustrate the Fe\,{\footnotesize II} subtraction procedure, we
plot three spectra for \rgb between 4300 and 5600 {\AA} in the rest frame of 
the source. The top one is the observed spectrum. The bottom one is the 
Fe\,{\footnotesize II} model derived from the I Zw 1 template of Boroson \& 
Green (\cite{borosong}) by folding it with a Gaussian of 980 km s$^{-1}$ FWHM 
and normalizing it to about 8\% of the flux of I Zw 1. The middle one is the 
Fe\,{\footnotesize II} subtracted spectrum shifted downward by an arbitrary 
amount for illustrative purposes.}
\label{opt}
\end{figure}

The spectrum of \rgb shown in Laurent-Muehleisen et al. (\cite{laurent98}) and
on which its NLS1 classification is based, did not extend far enough to the 
red to include H$\alpha$. We therefore obtained a second spectrum on Sept.\ 
24, 1998 at Lick observatory using the Kast double spectrograph. Two 20 minute 
exposures were taken to facilitate accurate cosmic ray rejection and the 
resulting averaged spectrum has a resolution of 3.8 {\AA} and a coverage from 
3410 to 8270 {\AA} (2900 to 7000 {\AA} in the object's rest frame). Weather 
was variable, but generally poor during the observations. Therefore the flux 
scale should be taken only as a rough indicator of the source's brightness. 
Atmospheric lines were removed using an observation of a spectrophotometric 
standard star. The observations of both source and standard star were done 
with the slit rotated to the parallactic angle. We did not correct for 
reddening since the measured H$\alpha$/H$\beta$ ratio is consistent with no 
reddening.

The optical spectrum of \rgb appears to be extremely blue (Fig.~\ref{optb}). 
Even when the contributions from the Balmer continuum, blended Fe emission and
possibly Mg\,{\footnotesize II} are taken into account, the continuum below 
5000{\AA} is still steeper than the limit for a bare accretion disk spectrum
($\alpha \approx -0.3$; e.g. Grupe et al. 1998). This may indicate the 
presence of additional line emission, perhaps from Fe~{\footnotesize I} 
(e.g. Kwan et al. 1995), or scattering of a blue intrinsic continuum . If we 
restrict our measurement of the continuum to the region redward of 5000{\AA}, 
we find that a power law with a slope\footnote{$f_{\nu} \propto \nu^{-\alpha}$}
of $\alpha = 1.3$ yields a good fit. For wavelengths blueward of 5000{\AA}, 
we formally get a slope of $\alpha = -3.1$. 

To our knowledge, this is bluest optical continuum of any NLS1 observed 
to date. It might be speculated that this peculiarity 
is related to the radio-loudness of RGB\,J0044+193, for example via optical
synchrotron emission from the putative radio jet. However, the spectral 
shape practically rules out 
a synchrotron contribution to the blue optical continuum. We further note 
that the optical continuum of the only other known radio-loud NLS1 
PKS 0558-504 is not as blue as that of \rgb (Remillard et al. 
\cite{remillard}).

The emission line spectrum confirms the original classification as a NLS1 
galaxy at redshift of z=0.181.  The broad permitted blend of 
Fe\,{\footnotesize II} emission commonly present in NLS1s severely contaminates
the spectrum, making an accurate measurement of the flux and width, in 
particular of the H$\beta$ and [O\,{\footnotesize III}] lines difficult.  In 
order to account for this, we employ the Fe\,{\footnotesize II} subtraction 
method introduced by Boroson \& Green (\cite{borosong}) and now commonly 
applied (e.g. Leighly \cite{leighly99}). The Fe\,{\footnotesize II} emission 
line 
spectrum from a high signal-to-noise optical spectrum of the prototype strong 
Fe emitter I Zw 1 is first convolved with a Gaussian and then scaled until the
width and intensity of the lines approximately match those seen in the \rgb 
spectrum. We find that a 980 km s$^{-1}$ FWHM Gaussian convolution with a 
normalization 8\% of the absolute flux of I Zw 1 provides the best fit (see 
Fig. \ref{opt}). This best-fitting Fe\,{\footnotesize II} model was then 
subtracted and the remaining emission lines examined.

H$\beta$ has a Lorentzian rather than a Gaussian profile, which is typical of 
the NLS1 class (Goncalves et al. 1998). A Lorentzian 
profile with a FWHM of $\sim$1330 km s$^{-1}$ yields an excellent fit. 
H$\alpha$ is blended with [N\,{\footnotesize II}], but also shows clear 
indications of a broad component.  Our best fit Lorentzian profile yields a 
FWHM of 1000 km s$^{-1}$. Considering the degree of blending near H$\alpha$, 
this is fully consistent with the dynamics exhibited by the H$\beta$ line.  
The forbidden [O\,{\footnotesize III}] lines are well fit by a Gaussian of 
FWHM $\sim$750 km s$^{-1}$.  This behavior is typical for NLS1s: the broad and
narrow components of the permitted lines cannot be clearly separated, but 
the FWHM is $<$2000 km s$^{-1}$, while the forbidden lines are well-fit by a 
single narrow component.  The final pieces of optical evidence, which support 
the classification of \rgb as a NLS1 are the intensity ratios of the high 
ionization lines ([O\,{\footnotesize III}], [N\,{\footnotesize II}]) to the 
low ionization lines (H$\beta$, H$\alpha$). Because the 
H$\alpha$/[N\,{\footnotesize II}] emission feature is strongly blended,
we cannot determine an accurate flux ratio, but we can confidently say that 
[N\,{\footnotesize II}]/H$\alpha$$<$0.4. For [O\,{\footnotesize III}]/H$\beta$ 
we derive a value of 0.2.  Both these measurements are consistent with the 
low values typical for other NLS1 galaxies (Osterbrock \& Pogge 
\cite{osterbrock}; Goodrich \cite{goodrich}), and significantly lower 
than those exhibited by quasars, where [O\,{\footnotesize III}]/H$\beta$$>$3.0 
and [N\,{\footnotesize II}]/H$\alpha$$>$0.75.

\section{The radio source associated with \rgb}

Applying the criterium of Kellerman et al. (\cite{kellermann}), who define 
R = 10 as the dividing line between radio-loud and radio-quiet 
quasars\footnote{R is defined as the K-corrected ratio of the 4.85\,GHz radio 
flux and the optical flux at 4400 {\AA}.}, the source is radio-loud, since 
the ratio of the total 4.85\,GHz radio flux from the 87GB survey 
($f_{\rm 4.85GHz}$ = 24 mJy) and the optical magnitude ($m_{\rm O} = 16.7$) 
is $R\approx 31$. The radio flux corresponds to a luminosity\footnote{$H_0$ 
= 50 km s$^{-1}$ Mpc$^{-1}$, $q_0 = 0.5$.} of $\log P = 24.5$ W\,Hz$^{-1}$. 
Thus, also using luminosity criteria (e.g. Joly 1991; Miller et al. 1993), 
\rgb should be classified as a radio-loud AGN. 

In general, NLS1 galaxies are considered to be radio-quiet AGN. Ulvestad et 
al. (\cite{ulvestad}) investigated the radio emission of 15 NLS1 galaxies 
with the VLA down to a flux limit of 0.25 mJy at 5\,GHz. They found 
luminosities (or upper limits) between $\log P = 20.5$ W Hz$^{-1}$ and
$\log P = 22.5$ W Hz$^{-1}$. Thus, \rgb is about two orders of magnitude 
brighter at 4.85\,GHz than any of the objects in the sample of Ulvestad et al. 
(\cite{ulvestad}). Comparing the radio-loudness R of \rgb to that of the 
Ulvestad et al. (\cite{ulvestad}) sources, we again find that it is a factor
of $\sim$6 higher than any of those. Even if we use the lower flux from the
VLA measurement (see below) R is higher by a factor of two. However, in this
case R$<$10 and hence \rgb is technically radio-quiet (but confer discussion 
below). There is only one other radio-loud NLS1 galaxy known up to now (PKS 
0558-504) and we compare it to \rgb in Sect.\,7.1. 

The radio luminosity of \rgb is much larger than usually detected in normal
galaxies or starbursts. Smith et al. (1998), for example, study 
the 20 most radio luminous starburst galaxies from the UGC sample. The 
average 4.85 GHz luminosity of this sample is  $\langle\log P\rangle\approx 
23$, and even the radio brightest object is more than a factor of five less 
luminous than RGB\,0044+193. Therefore, any significant contribution to the 
total radio emission other than non-thermal radiation from a radio jet seems 
rather unlikely. 

Interestingly, \rgb was detected with a 4.85\,GHz flux of only 7 mJy in the 
high resolution follow-up observation with the VLA (Laurent-Muehleisen et al. 
\cite{laurent97}). This indicates that the source is either extended or 
variable. The latter is supported by two additional pieces of evidence: 
firstly, the radio source appears unresolved on the VLA map and secondly, 
it is not detected in the NRAO/VLA Sky Survey (NVSS; Condon et al. 
\cite{condon98} ) at 1.4\,GHz, which is 
sensitive down to $\sim$ 2.5 mJy. Given that extended radio emission generally 
shows steep radio spectra, the 1.4\,GHz radio flux should have been higher 
than 7 mJy and hence easily detectable in the NVSS. Unfortunately, existing
radio data do not allow us to place limits on the likelihood for large
intrinsic radio variability in NLS1 galaxies as a class. Even in the case of 
\rgb the true amount of intrinsic variability is uncertain, since the sense 
of the differences between the 87GB and the VLA measurements is the same as 
if overresolution were a problem. Due to the compactness of the radio source
interstellar scintillation might also be relevant.   

If taken at face value, the data also imply an inverted spectrum between 
4.85 and 1.4 GHz, probably due to synchrotron self-absorption. This would 
be consistent with the observed unresolved compact radio core. Note that 
neither PKS 0558-504 nor the Ulvestad et al. (\cite{ulvestad}) sources show 
inverted spectra.

\section{The ASCA observation}

\rgb was observed with ASCA (Tanaka et al. \cite{tanaka}) in 1-CCD faint mode from 
January 9, 08:55:23 to January 10, 23:49:13 (UT) 1998.

The data were analysed using {\footnotesize FTOOLS 4.1}. The recommended 
standard 
screening criteria were applied to the data. For the GIS the minimum elevation
angle above the Earth's limb ({\tt ELV}) was chosen to be $5\degr$. In the 
case of the SIS we used {\tt ELV}$ >10\degr$. To avoid atmospheric 
contamination, data were only accepted in the SIS when the angle between the 
target and the bright earth ({\tt BR\_EARTH}) was greater than 
$20\degr$. Further, a minimum cut-off rigidity ({\tt COR}) of 6 GeV/c was 
applied for both, the SIS and the GIS. Data taken within 60 seconds after
passage of the day-night-terminator and the South Atlantic Anomaly (SAA) were 
not considered in the analysis. Periods of high background
were manually excluded from the data by checking the light curve of the
observation. The resulting effective exposures were 47.9 ksec for each GIS
and 45.1 and 46.4 ksec for SIS0 and SIS1, respectively.

Source counts were extracted from a circular region centered on the target 
with a radius of $6\arcmin$ for the GIS and $4\arcmin$ for the SIS.
We used the local background determined from the observation in the
analysis for both detectors. In particular, the GIS background was estimated 
from a source free region at the same off-axis angle as the source and with 
the same size as the source extraction region. In total, $\sim$2500 counts 
were detected from \rgb in each of the SIS detectors. Because of the steep
spectrum and the lower effective area, the number of photons detected with 
the GIS detectors was a about factor of two smaller.
 
All spectra were rebinned to have at least 20 photons in each energy channel. 
This allows the use of the $\chi^2$ technique to obtain the best fit values 
for the model spectra. We used the latest GIS redistribution matrices 
available (V4\_0) from the calibration database and created the SIS response 
matrices for our observation using {\footnotesize SISRMG}, which applies the
latest charge transfer inefficiency (CTI) table ({\tt sisph2pi\_110397.fits}).
The ancillary response files for all four detectors were generated using the 
{\footnotesize ASCAARF} program. 

Spectra were fitted in the energy range 0.8 to 8 keV for both GIS. For SIS0 
and SIS1 we used 0.5 to 7 keV and 0.7 to 7 keV, respectively. The upper energy 
boundaries are given by the maximum energy at which the source was detected 
in each instrument. The lower energy boundaries result from the calibration 
uncertainties of the detectors. In particular, SIS1 recently shows systematic 
residuals below 0.6 keV (Dotani et al. \cite{dotani}). Also, since the CCD 
temperature of SIS1 was higher than SIS0 during the observation, we decided 
to ignore the data below 0.7 keV for SIS1. An increasing RDD (Residual Dark 
Distribution), which is caused by radiation damage, has been reported for both 
CCDs. This effect cannot be corrected for with currently available software.
RDD degradation should be negligible for 1-CCD observations, however 
(Dotani et al. \cite{dotani}).

\section{Spectral analysis}

Since the spectral fitting for the individual detectors gives consistent
results, we only cite the parameters for all four detectors fitted 
simultaneously in the following.
The normalization of the GIS detectors was allowed to vary with respect
to the SIS, because of the known cross-calibration uncertainties. The
resulting differences in the normalization between the SIS and the GIS 
detectors are of the order of 5 to 10\%.

\begin{figure}
\resizebox{\hsize}{!}{\includegraphics{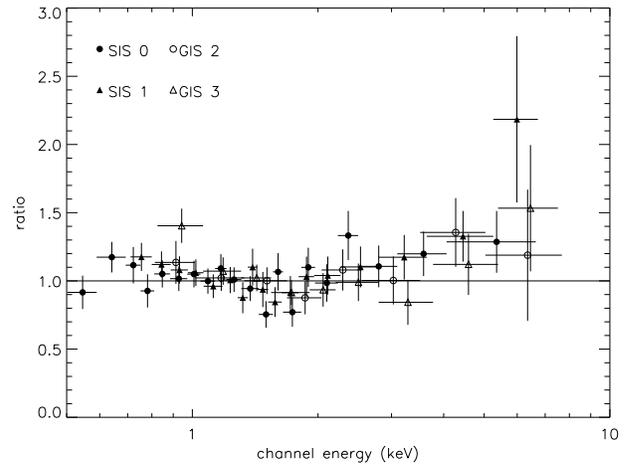}}
\caption{Ratio of the data to a simple power law model. The break in
the X-ray spectrum around 1.8 keV can clearly be seen.}
\label{pl}
\end{figure}

\begin{figure}
\resizebox{\hsize}{!}{\includegraphics{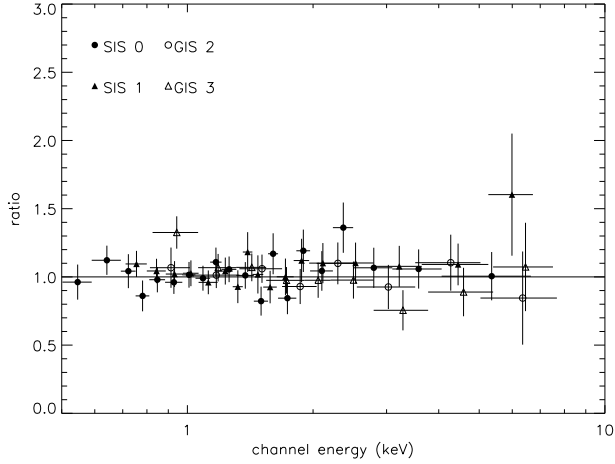}}
\caption{Ratio of the data to a model including a blackbody with T$= 0.19\pm 
0.02$ keV and a power law with a photon index of $\Gamma = 2.1$.}
\label{bb}
\end{figure}

We first investigated a simple power law model with the absorption
fixed to the Galactic value (N$_H = 3.55\times 10^{20}$ cm$^{-2}$).
The resulting best fit parameters are $\Gamma = 2.47\pm0.07$ and
$\chi^2$ = 405.5 (360 d.o.f.). The ratio of the data to this model
is shown in Fig.~\ref{pl}. It can clearly be seen that there is a
significant break in the spectrum of \rgb around 1.8 keV.

Next we fitted the spectrum with a broken power law model. To put
tighter constraints on the break energy and the low energy power law,
we first fitted a simple power law model to the data above 2 keV and
fixed the resulting high energy power law ($\Gamma\approx 2.1$) in the
broken power law fit. This gives $\Gamma_{\rm low} = 2.69\pm0.11$, 
$E_{\rm break} = 1.79^{+0.35}_{-0.20}$ keV and $\chi^2$ = 386.5 (359 d.o.f.). 
The difference in $\chi^2$ is significant at more than 99.9\% confidence
according to a F--test. 

An even better fit results when a blackbody model superimposed on a simple 
power law is applied instead of a broken power law. Using {\tt ZBBODY} we get 
(again for fixed Galactic absorption) T = 0.19$\pm$0.02 keV for a power law
photon index of $\Gamma = 2.1$ (fixed). The resulting $\chi^2$ is 382.6 for
359 degrees of freedom. The resulting ratio of data to model is shown in 
Fig.~\ref{bb}. For this model we get a 2--10 keV flux of $(7.9\pm0.3)\times 
10^{-13}$ erg\,cm$^{-2}$\,s$^{-1}$, which gives a rest frame 2--10 keV 
luminosity of $(1.35\pm0.05)\times 10^{44}$ erg s$^{-1}$. The corresponding 
soft X-ray (0.5-2 keV) luminosity is $1.9\times 10^{44}$ erg s$^{-1}$.

Alternatively, if we use a thermal bremsstrahlung model ({\tt ZBREMSS}) 
instead of a blackbody, we also get an acceptable fit and the best fitting 
parameters are T = 0.45$^{+0.10}_{-0.07}$ keV, $\Gamma = 2.1$ (fixed) and 
$\chi^2 = 385.9$ (359 d.o.f.). We note that leaving the photon index as a 
free parameter in these models gives values that are consistent with the 
photon index determined at higher energies.

We do not detect any indication of absorption edges due to O\,{\footnotesize 
\sc VII} and O\,{\footnotesize \sc VIII} and the upper limits on the optical 
depths are $\tau$({\footnotesize O\,VII})$< 0.50$ and $\tau$({\footnotesize 
\rm O\,VIII})$< 0.54$. We also searched for a K$\alpha$ 
emission line from neutral or ionized iron, but could not find any. Due to the 
low photon statistics at high energies the upper limits on the equivalent 
widths are not very stringent: 250 eV and 400 eV for a narrow emission line at
6.4 and 6.7 keV, respectively.  

\section{The light curve}

\begin{figure}
\resizebox{\hsize}{!}{\includegraphics{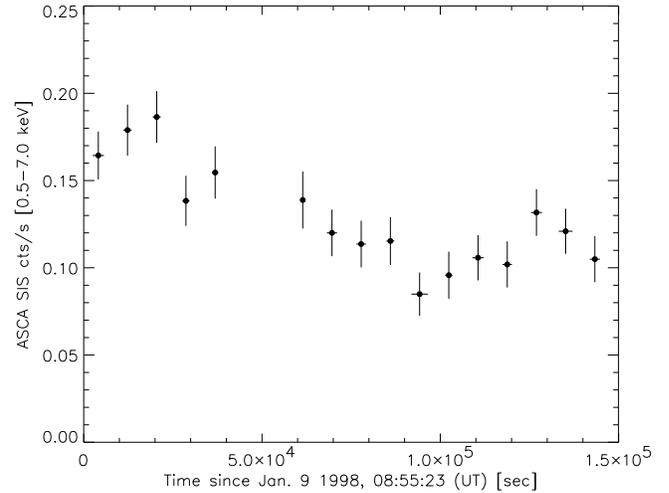}}
\caption{X-ray light curve of \rgb for the combined SIS detectors 
in the 0.5 to 7 keV range.}
\label{light}
\end{figure}

The background subtracted light curve of \rgb in the 0.5--7.0 keV 
energy range during the ASCA observation is shown in Fig.~\ref{light}. The 
data were binned in such a way that each data point corresponds to one
orbit of the satellite. For clarity, only the combined SIS0/SIS1 light curve 
is shown, but we note that the same qualitative behavior is also seen in the 
GIS detectors.

The source count rate was clearly higher at the beginning of the observation. 
It decreased by a factor of two within one day. There is also evidence for
significant variability on much shorter time scales, albeit with smaller
amplitudes.

In the ROSAT All-Sky Survey \rgb was detected with a PSPC count rate of
$0.17\pm0.02$ counts s$^{-1}$. Using a power law model with $\Gamma = 2.7$,
this results in an unabsorbed 0.1-2.4 keV flux of $(4.8\pm0.6)\times 
10^{-12}$ erg cm$^{-2}$ s$^{-1}$. This value is consistent with 
the corresponding 
ASCA flux for the broken power law model in the same energy range ($(5.2\pm0.2)
\times 10^{-12}$ erg cm$^{-2}$ s$^{-1}$). 
 
\begin{figure}
\psfig{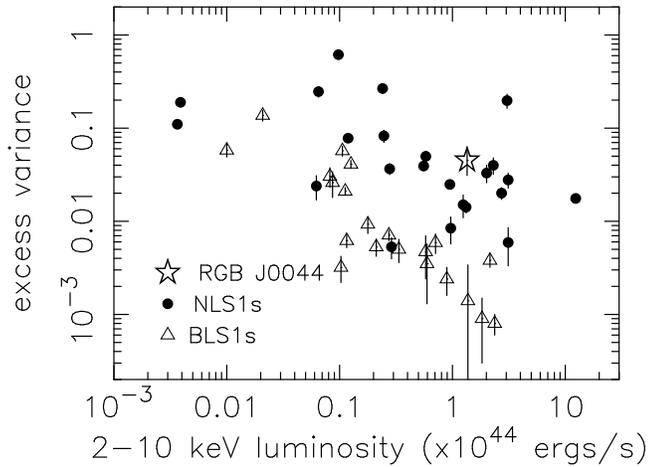}
%\resizebox{\hsize}{!}{\includegraphics{new_fig5.ps,angle=-90}}
\caption{Excess variance vs. 2-10 keV luminosity for \rgb compared with 
broad-line and narrow-line Seyfert 1 galaxies; figure taken from Leighly 
(\cite{leighly99}). At a given X-ray luminosity NLS1 galaxies clearly show 
more rapid variability than broad line objects (see also Leighly 
\cite{leighly98}). \rgb falls into the region of NLS1 galaxies, thus 
providing independent evidence for the NLS1 nature of RGB\,J0044+193.}
\label{excess}
\end{figure}

A common measure of AGN X-ray variability is the excess variance $\sigma_{e}$
(Nandra et al. \cite{nandra}; Leighly \cite{leighly99}). Basically it 
measures the variance of the light curve minus the variance due to the 
measurement errors. For exposures of 
similar duration it is also a measure of the variability time scale. In the 
case of \rgb we calculate $\sigma_e = 0.045\pm0.014$, which is 
significantly higher than the values found for broad-line Seyfert 1 galaxies 
of similar X-ray luminosity, but consistent with the values found for other
NLS1s (Leighly \cite{leighly99}). This is illustrated
in Fig.~\ref{excess} (adapted from from Fig. 3 of Leighly \cite{leighly99}), 
which shows the excess variance of \rgb in the total band in comparison to 
broad-line and narrow-line Seyfert 1 galaxies.

Given the evidence for a soft component in the X-ray spectrum of \rgb,
we also calculated the excess variance above and below 1.8 keV. We get 
0.040$\pm0.014$ at low energies and 0.093$\pm$0.031 at higher energies. 
Although there seems to be a tendency for more pronounced variability at 
higher energies, we do not consider the difference in the excess variances 
of the soft and the hard band as significant.

\section{Comparison to other NLS1 galaxies}

\subsection{The second radio-loud NLS1 PKS 0558-504}

The only other known radio-loud NLS1 galaxy is PKS 0558-504. It is at 
a redshift of $z = 0.137$, has a optical magnitude of $m_{\rm B} = 14.97$ 
and a 4.85 GHz radio flux of 113 mJy. This gives a radio-loudness of 
$R\approx27$ and a radio luminosity of $\log P = 24.9$. These parameters are
very similar to those of \rgb.

In X-rays, PKS 0558-504 is an order of magnitude more luminous than \rgb.
Leighly (1999) gives a 0.5--2 keV luminosity of $\approx1.7\times 
10^{45}$ erg $^{-1}$ (as compared to $1.9\times 10^{44}$ erg s$^{-1}$ for
RGB\,J0044+193). Despite this huge difference in luminosity, the ASCA spectra
of the two radio-loud NLS1 galaxies are remarkably similar. Leighly (1999)
finds a slightly higher temperature (kT $\approx$ 0.25 keV) for the blackbody 
model and a slightly steeper hard photon index ($\Gamma \approx 2.25$). 
Interestingly, the 
spectrum of PKS 0558-504 also shows marginal evidence for an emission line 
around 6.7 keV with an equivalent width of about 100 eV, which is well 
below the upper limit for RGB\,J0044+193, however.   

\subsection{Radio-quiet NLS1s}

It was shown by Brandt et al. (\cite{brandt}), that the hard (2--10 keV) 
X-ray spectrum of NLS1 galaxies is steeper on average than that of broad line 
Seyfert 1 galaxies, although the scatter in spectral indices seems to be 
smaller compared to ROSAT PSPC measurements (Boller et al. \cite{boller96}).
Our result for the photon index of \rgb at energies above 2 keV 
($\Gamma\approx 2.1$) is close to the average value for NLS1 (Brandt et al. 
\cite{brandt}; Leighly \cite{leighly99}). Although the hard X-ray spectra of 
NLS1 galaxies
are significantly steeper than those of normal Seyfert 1 galaxies, they 
are nevertheless much flatter than the soft X-ray spectra of NLS1 galaxies
as measured with ROSAT. This clearly indicates a break in the X-ray spectrum. 
In Sect. 5 we have presented unambiguous evidence for the presence of 
such a break in \rgb and the ASCA photon index below the break energy 
($\Gamma_{\rm low}\approx 2.7$) is consistent with the ROSAT spectral indices 
of NLS1 galaxies (Boller et al. \cite{boller96}). Similar energy breaks are 
found in a systematic study of the ASCA spectra of NLS1 galaxies (Leighly
\cite{leighly99}). We conclude that the X-ray 
spectral properties of \rgb do not appear to be different from radio-quiet 
NLS1 galaxies. 

For radio-loud AGN a correlation between the core-dominance and the X-ray
spectral index has been claimed in the sense that the most core-dominated
sources have the flattest X-ray spectra (e.g. Kembhavi \cite{kembhavi}; 
Brinkmann et al. \cite{bys}). This result is usually interpreted in terms 
of a flat inverse Compton component in the X-ray spectrum, which is related
to the relativistic radio jet and which dominates the emission spectrum for 
objects oriented close to the line of sight. The seed photons for this 
process are either the synchrotron photons generated in the radio jet itself
or external photons from the accretion disc or the emission line regions.
In this scenario, the apparent absence of a flat X-ray component in \rgb 
might indicate that the radio jet is oriented at a rather large angle to 
the line of sight, which would argue against a pole-on geometry at least 
for this source. However, there are several caveats. First of all, nothing 
is known about the physical parameters related to the putative radio jet 
in \rgb or NLS1s in general. They might be different from those in radio-loud 
quasars. Secondly, \rgb is barely radio-loud. Therefore any inverse Compton 
component (e.g. synchrotron self-Compton) might simply be too weak to be 
detected in our ASCA observation.

Fig.~\ref{excess} illustrates that \rgb is significantly more variable than 
Seyfert 1 
galaxies of the same X-ray luminosity. The excess variance is more than one 
order of magnitude larger.  In fact, \rgb clearly falls into the 
region populated by the bulk of NLS1 galaxies (Leighly \cite{leighly99}). 
Also in terms of their variability properties, there seems to be no 
difference between \rgb and radio-quiet NLS1 galaxies.

\section{Summary}

We presented radio, optical and X-ray data of the peculiar radio-loud NLS1 
galaxy RGB\,J0044+193. The results can be summarized as follows:
\begin{enumerate}
\item The radio observations indicate significant variability in the flux 
from the compact radio source associated with \rgb and evidence for
an inverted radio spectrum.
\item The optical spectrum clearly confirms the NLS1 classification for
RGB\,J0044+193. In addition, the optical continuum appears to be extremely 
blue. This might indicate additional line emission or scattering of a blue
intrinsic continuum. 
\item The X-ray spectrum obtained with the ASCA satellite shows a break near 
2 keV. At higher energies it is best described by a power law model with a 
photon index of $\Gamma = 2.1$. The soft component can be modeled with either 
a $\sim0.2$ keV blackbody or a steeper power law ($\Gamma\approx 2.7$).
\item The ASCA SIS count rate decreases by a factor of two within one day
and there is evidence for variability with lower amplitude on much smaller 
time scales. The variability behavior is characterized by an excess variance 
of 0.045$\pm$0.014.
\end{enumerate}
Both the spectrum and the variability properties are typical for NLS1 
galaxies, which provides independent evidence for the classification of 
RGB\,J0044+193. We do not detect any differences between \rgb and the bulk 
of radio-quiet NLS1 galaxies. In particular, we find no evidence for a flat
X-ray component due to inverse Compton emission related to the putative 
non-thermal radio emission of RGB\,J0044+193. This does not necessarily argue
against a pole-on orientation for RGB\,J0044+193, however. It might well be
possible that the physical parameters of the putative radio jet in this
source render an inverse Compton component undectable within our ASCA 
observation. 

\begin{acknowledgements}
JS acknowledges financial support from the RIKEN--MPG exchange program. 
JS also thanks his colleagues from the Cosmic Radiation Laboratory at RIKEN 
for hospitality and support during a stay at the institute, where part of 
this work was done. We wish to thank Michael Brotherton for useful discussions
on iron template line fitting. KML gratefully acknowledges support through 
NAG5-7261 ({\it ASCA}). SALM acknowledges support from the Department 
of Energy at the Lawrence Livermore National Laboratory under contract 
W-7405-ENG-48.  This research has made use of the ASCA IDL Analysis System 
developed by Tahir Yaqoob, Peter Serlemitsos and Andy Ptak and of the 
NASA/IPAC Extragalactic Data Base (NED), which is operated by the Jet 
Propulsion Laboratory, California Institute of Technology, under contract 
with the National Aeronautics and Space Administration.
\end{acknowledgements}


\begin{thebibliography}{}
\bibitem[1995]{baker} Baker J.C., Hunstead R.W., 1995, ApJ 452, L95
\bibitem[1996]{boller96} Boller Th., Brandt W.N., Fink H.H., 1996, A\&A 305, 53
\bibitem[1997]{boller97} Boller Th., Brandt W.N., Fabian A.C., Fink H.H., 1997,
MNRAS 289, 393
\bibitem[1992]{boroson} Boroson T.A., 1992, ApJ 399, L15
\bibitem[1992]{borosong} Boroson T.A., Green R.F., 1992, ApJS 80, 109
\bibitem[1997]{brandt} Brandt W.N., Mathur S., Elvis M., 1997, MNRAS 285, 
25{\footnotesize P}
\bibitem[1995]{brinkmann95} Brinkmann W., Siebert J., Reich W., et al., 1995, 
A\&AS 109, 147
\bibitem[1997a]{brinkmann97} Brinkmann W., Siebert J., Feigelson E.D., et al., 
1997a, A\&A 323, 739 
\bibitem[1997b]{bys} Brinkmann W., Yuan W., Siebert J., 1997b, A\&A 319, 413
\bibitem[1989]{condon} Condon J.J., Broderick J.J., Seielstad G.A., 1989, AJ 
97, 1064
\bibitem[1998]{condon98} Condon J.J., Cotton W.D., Greisen E.W., et al., 
1998, AJ 115, 1693
\bibitem[1997]{dotani} Dotani T., Yamashita A., Ezuka H., et al., 1997, ASCA News No. 5 
\bibitem[1998]{goncalves} Goncalves A.C., V\'eron P., V\'eron-Cetty M.-P., 
1998, in: Structure and Kinematics of the Quasar Broad Line Regions, 
Gaskell C.M., Brandt W.N., Dietrich M., Dultzin-Hacyan D., Eracleus M. (eds.),
in press
\bibitem[1989]{goodrich} Goodrich R.W., 1989, ApJ 342, 224
\bibitem[1998]{grupe} Grupe D., Beuermann K., Thomas H.-C., Mannheim K., 
Fink H.H., 1998, A\&A 330, 25
\bibitem[1993]{haardt} Haardt F., Maraschi L., 1993, ApJ 413, 507
\bibitem[1998]{iwasawa} Iwasawa K., Brandt W.N., Fabian A.C., 1998, MNRAS 293,
251  
\bibitem[1991]{joly} Joly M., 1991, A\&A 242, 49
\bibitem[1989]{kellermann} Kellermann K.I., Sramek R., Schmidt M., Shaffer 
D.B., Green R., 1989, AJ 98, 1195
\bibitem[1993]{kembhavi} Kembhavi A., 1993, MNRAS 264, 683
\bibitem[1995]{kwan} Kwan J., Cheng F., Fang L., Zheng W., Ge J., 1995, 
ApJ 440, 628 
\bibitem[1997]{laurent97} Laurent-Muehleisen S.A., Kollgaard R.I., Ryan P.J., 
et al., 1997, A\&AS 122, 235
\bibitem[1998]{laurent98} Laurent-Muehleisen S.A., Kollgaard R.I., Ciardullo 
R.B., et al., 1998, ApJS 118, 127
\bibitem[1998]{leighly98} Leighly K.M., 1998, in: Proc. ``Accretion Processes 
in Astrophysical Systems: Some Like it Hot!'', Holt S.S, Kallman T.R. (eds.), 
Woodbury, New York, p. 199
\bibitem[1999]{leighly99} Leighly K.M., 1999, ApJ submitted
\bibitem[1997]{leighly97} Leighly K.M., Mushotzky R.F., Nandra K., Forster 
K., 1997, ApJ 489, L25
\bibitem[1988]{madau} Madau P., 1988, ApJ 327, 116
\bibitem[1998]{miller} Miller P., Rawlings S., Saunders R., 1993, MNRAS 263, 
425
\bibitem[1997]{nandra} Nandra K., George I.M., Mushotzky R.F., Turner T.J., 
Yaqoob T., 1997, ApJ 476, 70
\bibitem[1985]{osterbrock} Osterbrock D.E., Pogge R.W., 1985, ApJ 297, 166
\bibitem[1995]{pounds} Pounds K.A., Done C., Osborne J.P., 1995, MNRAS 277,
5{\footnotesize P} 
\bibitem[1986]{remillard} Remillard R.A., Bradt H.V., Buckley D.A.H., et al., 
1986, ApJ 301, 742
\bibitem[1998]{smith} Smith D.A., Herter T., Haynes M.P.,1998, ApJ 494, 150
\bibitem[1994]{tanaka} Tanaka Y., Inoue H., Holt S.S., 1994, PASJ 46, L37
\bibitem[1995]{ulvestad} Ulvestad J.S., Antonucci R.R.J., Goodrich R.W., 1995,
 AJ 109, 81
\bibitem[1995]{wilson} Wilson, A.S., 1995, in: Proceedings of 
the Oxford Torus Workshop, Ward M.J. (ed.), p.55 
\end{thebibliography}
\end{document}